\documentclass[twocolumn,amsmath,amssymb,aps,pra]{revtex4-1}
\usepackage{graphicx}
\usepackage{bm}

\usepackage{feynmp-auto}
\usepackage{amsmath}
\usepackage{amsfonts}

\usepackage[colorlinks=true, allcolors=blue]{hyperref}
\usepackage{nameref}
\usepackage{cleveref}

\newcommand{\ii}{\mathrm{i}} 

\begin{document}

\title{Long-range magnetic interaction within quantum electrodynamics formalism}
\author{D. Solovyev$^{1,2}$}
\email[E-mail:]{d.solovyev@spbu.ru}

\author{T. Zalialiutdinov$^{1,2}$}

\author{A. Anikin$^{1,3}$}
\author{A. Bobylev$^{1,2}$}
\author{P. Kvasov$^1$}
\author{J. J. Lopez-Rodriguez$^1$}
\author{A. Moshkin$^{1}$}
\author{D. Zinenko$^{1}$}
\author{D. Glazov$^{4,2}$}
\affiliation{ 
$^1$ Department of Physics, St. Petersburg State University, Petrodvorets, Oulianovskaya 1, 198504, St. Petersburg, Russia}
\affiliation{ $^2$ Petersburg Nuclear Physics Institute named after B.P. Konstantinov of National Research Centre 'Kurchatov Institut', St. Petersburg, Gatchina, 188300, Russia}
\affiliation{ $^3$ D. I. Mendeleev Institute for Metrology, St. Petersburg, 190005, Russia}
\affiliation{ $^4$ School of Physics and Engineering, ITMO University, Kronverkskiy pr. 49, 197101 St. Petersburg, Russia
}

\date{\today}

\begin{abstract}
Within the framework of quantum electrodynamics, the interaction between two atoms at large distances is analyzed. Using the S‑matrix formalism, an expression for the magnetic interaction potential is derived, which agrees with the well-known result of classical electrodynamics. However, quantum electrodynamics goes beyond this conventional result and allows one to treat a wide range of problems related to the structure of atomic energy levels. In particular, it is shown that the asymptotic behavior of the interaction potential can deviate from the classical prediction, depending on the atomic states involved. As an example, dispersion coefficients are calculated for the s-states of hydrogen atoms, where the long-range potential reduces to a spin–spin interaction. The results obtained open up the possibility of a straightforward comparative analysis of long-range interaction potentials between atoms of matter and antimatter. The applicability of this approach is demonstrated for the hydrogen–antihydrogen system.
\end{abstract}

\maketitle

\section{Introduction}

The investigation of long-range interactions between atoms and atomic systems, including their applications within solid-state theory and molecular physics, constitutes a well-established field of research. Beginning with the foundational papers \cite{London1930,CasimirPolder:1948}, the long-range interaction between two stationary neutral atoms is generally described using fourth-order perturbation theory. The reason is that in second order, the relevant dipole matrix elements vanish (they are diagonal for an unchanged internal system state). Consequently, at large distances, the dipole-dipole interaction arises via atomic polarizabilities and, to leading order, is given by the London dispersion formula. Nowadays, the derivation of the classical van der Waals interaction can be found in standard textbooks on quantum mechanics (QM) or quantum electrodynamics (QED) \cite{Landau,Berest,Kaplan2006Intermolecular}.

Research in this area spans a wide range of topics across various branches of physics \cite{Kaplan2006Intermolecular,Israelachvili2011}. The advancement of quantum electrodynamic theory, in particular, relates not only to the most rigorous derivation of van der Waals interactions between atomic systems or their interactions with various surfaces \cite{Levitov_1989,PhysRevB.74.205413} but also to a more detailed account of atomic level structure \cite{PhysRevLett.118.123001,atoms5040048,atoms10010006}. This enables the detection of significant shifts in atomic energy levels in precision spectroscopic measurements \cite{PhysRevA.95.022703,PhysRevA.95.022704}. 

Interest extends beyond dispersion forces in isolated systems to include those under external influences, for example, from thermal radiation \cite{Lifshitz,Dzyaloshinskii1961}. Dipole-dipole interaction at large distances induced by heated environment continues to be an active area of research, see, e.g., \cite{PasseratdeSilans2014,Berman:2014,Safari:2015,Fujii:2022}. Very recently, long-range potentials induced by thermal Planck radiation were considered in the paper \cite{ZS-LR}. In particular, based on a strict QED approach, well-known zero-temperature expressions were obtained, that are usually phenomenologically generalized to the case of finite temperatures of the surrounding environment. The application of QED theory at finite temperature for bound states has shown that within this methodology, thermal induced interactions exhibit different asymptotic behavior at large distances compared to the phenomenological approach, see results and discussion in \cite{ZS-LR}. Based on the foregoing, it can be pointed out that the application of QED theory, including at finite temperatures, still remains an open question in the research of long-range potentials.

A significant area of research involves determining the long-range magnetic dipole-dipole interaction. In classical electrodynamics, the corresponding expression is well established and can be reduced to a spin-spin ($J$-coupling) interaction. Although the magnetic dipole-dipole interaction potential finds its primary application in solid-state physics \cite{Kittel2004,Kaplan2006Intermolecular}, we omit a detailed discussion of this vast field and its extensive literature for the sake of brevity. The primary goal of this work is to determine the magnetic coupling at large interatomic distances. To this end, we derive it from the first principles of quantum electrodynamics, carrying out analytical calculations under the assumption that the system state remains unchanged. For this purpose we employ the $S$-matrix formalism and consider one-photon exchange between two one-electron atoms, $A$ and $B$. At large fixed distances, where the dominant electric dipole-dipole term is zero, we demonstrate how the formalism naturally reveals the magnetic dipole-dipole interaction in the one-photon exchange picture.

This work is organized as follows. Section~\ref{section1} discusses the long-range interaction potential between two atoms within the $S$-matrix formalism for the case of one-photon exchange. A modification of the derived expression for finite temperatures is provided at the end of this section. Details of the numerical calculations and the accompanying analysis of the results are presented in Section~\ref{RD}, followed by Section~\ref{concl}, which contains the main conclusions of the study.

The paper employs relativistic units (r.u.), where $\hbar=c=m=1$ ($\hbar$ is Planck's constant, $m$ is the electron mass, and $c$ is the speed of light), and the fine structure constant is expressed in terms of the electron charge as $\alpha = e^2$. Boltzmann's constant in these units is $k_{\rm B}=m\alpha^2 k_{\rm B}^{\mathrm {a.u.}}$, where $k_{\rm B}^{\mathrm{a.u.}} = 3.16681\times 10^{-6}$ in atomic units. Four-dimensional coordinates are denoted by Latin letters, e.g., $x=(t,\bm{x})$, where the zeroth component corresponds to the time coordinate and the spatial components are set in bold.


\section{Long-range interaction within one-photon exchange approach}
\label{section1}

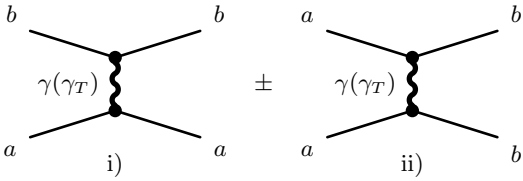
\begin{figure}[h!]
\begin{fmffile}{1ph}
\[
\vcenter{\hbox{
  \begin{fmfgraph*}(80,40)
    \fmfleft{i1,i2}
    \fmfright{o1,o2}
    \fmflabel{\(a\)}{i1}
    \fmflabel{\(b\)}{i2}
    \fmflabel{\(a\)}{o1}
    \fmflabel{\(b\)}{o2}
    \fmf{plain}{i1,v1,o1}
    \fmf{plain}{i2,v2,o2}
    \fmf{photon,width=2pt,label=\(\gamma(\gamma_T)\)}{v1,v2}
    \fmfdot{v1,v2}
    \fmfv{label=\text{i)},label.angle=90,label.dist=-25pt}{v1}
  \end{fmfgraph*}
}}
    \quad {\huge \pm} \quad
\vcenter{\hbox{
    \begin{fmfgraph*}(80,40)
        \fmfleft{i3,i4}
        \fmfright{o3,o4}
        \fmflabel{\(a\)}{i3}
        \fmflabel{\(a\)}{i4}
        \fmflabel{\(b\)}{o3}
        \fmflabel{\(b\)}{o4}
        \fmf{plain}{i3,v3,o3}
        \fmf{plain}{i4,v4,o4}
        \fmf{photon,width=2pt,label=\(\gamma(\gamma_T)\)}{v3,v4}
        \fmfdot{v3,v4}
        \fmfv{label=\text{ii)},label.angle=90,label.dist=-25pt}{v3}
  \end{fmfgraph*}
}}
\]
\end{fmffile}
\caption{Feynman diagrams depicting one-photon exchange between two atoms. The atomic states are labeled as $a$ and $b$ for atoms $A$ and $B$, respectively. For identical atoms, the exchange diagram should be taken into account: i) direct diagram, ii) exchange diagram. The wavy line represents a photon, whereas the electronic states are indicated by solid lines. The photon line notation $\gamma$ refers to the zero-temperature case, while $\gamma _{T}$ in parentheses corresponds to the case of finite temperatures. The Furry picture is implied.}
\label{fig:1}
\end{figure}

Within the framework of QED perturbation theory, the interaction between two atoms (hereafter denoted as $A$ and $B$) is given in the lowest order by one-photon exchange. To account for the symmetry or antisymmetry of the wavefunction for identical atoms, both direct and exchange diagrams should be included \cite{Chibisov}, as shown by Feynman diagrams in Fig.~\ref{fig:1}.


Due to the equivalence of identical atoms, states cannot be attributed to a particular atom ($A$ or $B$). This necessitates considering permutation symmetry, accounting for the appearance of graph ii). In the general case, this result is derived by constructing a system state vector that is either symmetric or antisymmetric under the permutation of constituent wavefunctions: $|\psi\rangle=(|\psi_A\psi_B\rangle \pm |\psi_B\psi_A\rangle)/\sqrt{2}$ (the sum or difference of the diagrams shown in Fig.~\ref{fig:1}). Such a state describes a scattering process in which two atoms can "interchange"\, their quantum state \cite{Chibisov,PhysRevLett.118.123001,PhysRevA.95.022703}. Consequently, the Van der Waals coefficient (mediated by the two-photon exchange diagrams), accounting for the symmetry of the state, was found as $C_{6}=D_{6}\pm M_{6}$, see \cite{PhysRevA.95.022703} for details.

Considering the interaction between two one-electron atoms (generalization of the theory to many-electron atomic systems can be performed, for example, using \cite{lindgren}), the $S$-matrix element can be written as follows.
\begin{eqnarray}
    \label{1}
    S_{AB}^{(2)}=(-\ii e)^2\int dx_1 dx_2 \overline{\psi}^{A}_{a}(x_1)\overline{\psi}^{B}_{b}(x_2)
    \\
    \nonumber
    \times
    \gamma^{\mu}_{A} D_{\mu\nu}(x_1,x_2)\gamma^{\nu}_B\psi^{A}_{a'}(x_1)\psi^{B}_{b'}(x_2),
\end{eqnarray}
where $\psi_{a}^{A}(x)=e^{-\ii\varepsilon_{a}^{A}t}\psi(\bm{x})$ is the solution of Dirac equation for bound electron in the state $a$ of the atom $A$, $\overline{\psi}_{a} = \psi_{a}^{+} \gamma_0$ is the Dirac conjugated wave function with $\psi_{a}^{+}$ being its Hermitian conjugate, $\gamma^{\mu}_{A} = (\gamma_0, \bm{\gamma})$ are the Dirac matrices (indexes $A$ and $B$ refers to the corresponding atoms), $\varepsilon_n$ is the Dirac energy. The components of the photon propagator $D_{\mu\nu}$ can be expressed as the sum of two contributions: the zero-temperature part $D^{0}_{\mu\nu}$ and the thermal one $D^{\beta}_{\mu\nu}$, which accounts for the Planck frequency distribution associated with photons in the thermal reservoir \cite{DHR,Don,SLP-QED}. The latter allows us to investigate and elucidate the influence of blackbody radiation on the interaction of two atoms separated by a distance $R$.


According to \cite{DonH}, the photon propagator in the Feynman (F) gauge and momentum space reads as
\begin{eqnarray}
    \label{2}
    D_{\mu\nu}^{\mathrm{DH}}(k)=D^{0, F}_{\mu\nu} +D^{\beta, F}_{\mu\nu} =
    \\
    \nonumber 
    -4\pi g_{\mu\nu}
    \left[\frac{\ii}{k^2+i0} + 2\pi \delta (k^2)n_{\beta}(|\bm{k}|) \right],
\end{eqnarray}
where $g_{\mu\nu}$ is the metric tensor of Minkowski space, $k=(k_{0},\bm{k})$ is the four-dimensional momentum, $k^2 = k_{0}^2 - \bm{k}^2$,
and $n_{\beta} $ is defined as follows
\begin{eqnarray}
    \label{3}
    n_{\beta}(\omega)=\frac{1}{e^{\beta\omega}-1}.
\end{eqnarray}
Here $\beta=(k_{\rm B}T)^{-1}$, $k_{\rm B}$ is the Boltzmann constant in relativistic units and $T$ is the temperature in Kelvin. In the coordinate space the photon propagator can be found by four-dimensional Fourier transform:
\begin{eqnarray}
    \label{4}
    D_{\mu\nu}^{\mathrm{DH}}(x_{1},x_{2})= -4\pi g_{\mu\nu}\int\frac{d^4k}{(2\pi)^4}e^{-\ii k(x_1-x_2)}    
    \\
    \nonumber
    \times
    \left[\frac{\ii}{k^2+i0} + 2\pi \delta (k^2)n_{\beta}(|\bm{k}|) \right].
\end{eqnarray}

Hereafter, it is convenient to use the temporal gauge (also known as the Weyl gauge) \cite{Berest}. 
Then the components of zero-temperature part of photon propagator are given as
\begin{eqnarray}
    D^{0}_{00}(k) = D^{0}_{0i}(k) = 0,
    \\
    \nonumber
    D^{0}_{ij}(k)=\frac{4\pi \ii}{k^2}\left(\delta_{ij}-\frac{k_{i}k_{j}}{k_{0}^2} \right).
\end{eqnarray}
For the  finite temperature part, see \cite{Escobedo2008,Escobedo2010}, we have
\begin{eqnarray}
    D^{\beta}_{00}(k) = D^{\beta}_{0i}(k) = 0,
    \\
    \nonumber
    D^{\beta}_{ij}(k)= 8\pi^2 \delta(k^2)\left(\delta_{ij}-\frac{k_{i}k_{j}}{k_{0}^2} \right)n_{\beta}(|k_0|).
\end{eqnarray}
In the above expressions, the indexes $i,\,j=1,\,2,\,3$ specify the transversal part of the photon propagator. 

The evaluation of corresponding coordinate representation of zero-temperature and thermal propagators in temporal gauge was presented in \cite{ZS-LR} with the result:
\begin{eqnarray}
    \label{7}
    D^{0,\beta}_{ij}(x_1,x_2)=
    \frac{\ii}{2\pi} \int\limits_{-\infty}^{\infty} dk_{0}e^{-\ii k_{0}(t_1-t_2)} D^{0,\beta}_{ij}(k_{0},r_{12}).
\end{eqnarray}
Here $r_{12}=|\bm{r}_{1}-\bm{r}_{2}|$ is the inter-electronic distance and
\begin{eqnarray}
    \label{8}
    D^{0}_{ij}(k_{0},r_{12})     =
    \left(\delta_{ij}+ \frac{\nabla_i \nabla_j}{k_0^2} 
    \right)
    \left\lbrace    -
    \frac{e^{\ii |k_0|r_{12}}}{r_{12}} \right\rbrace,
\\
\label{9}
    D^{\beta}_{ij}(k_{0},r_{12})
    =
   \left(\delta_{ij}+ \frac{\nabla_i \nabla_j}{k_0^2} 
    \right)
    \\
    \nonumber
    \times
    \left\lbrace
    -\frac{e^{\ii |k_{0}|r_{12}}}{r_{12}} 
    + \frac{e^{-\ii |k_{0}|r_{12}}}{r_{12}}
    \right\rbrace
    n_{\beta}(|k_{0}|).\qquad
\end{eqnarray}
The gradient operator, represented by the spatial component $\nabla_i$, acts on one of the variables $1$ or $2$ (otherwise, a minus sign would appear in the parentheses).

\begin{figure}
\centering
\begin{fmffile}{vector_diagram}
\begin{fmfgraph*}(130,110)
    \fmfset{arrow_len}{3mm}
    \fmfset{arrow_ang}{15}
    \fmfset{curly_len}{2mm}

    \fmfleft{i1,i2}
    \fmfright{o1,o2}

    \fmfforce{(0.2w,0.3h)}{v1}
    \fmfforce{(0.8w,0.2h)}{v2}
    \fmfforce{(0.9w,0.8h)}{v3}
    \fmfforce{(0.1w,0.7h)}{v4}

    \fmf{fermion,label=$\bm{R}$,label.side=right}{v1,v2}
    \fmf{fermion,label=$\bm{r}_B$,label.side=right}{v2,v3}
    \fmf{fermion,label=$\bm{r}_{12}$,label.side=left}{v4,v3}
    \fmf{fermion,label=$\bm{r}_A$,label.side=left}{v1,v4}

    \fmf{dashes,label.side=right,tension=0}{v1,v3}
    \fmf{dashes,label.side=right,tension=0}{v2,v4}

    \fmflabel{$Z_A e$}{v1}
    \fmflabel{$Z_B e$}{v2}
    \fmflabel{$-e_B$}{v3}
    \fmflabel{$-e_A$}{v4}    
\end{fmfgraph*}
\end{fmffile}
\caption{A diagram of radius vectors for two one-electron atoms at an internuclear distance $R$. The vector $\bm{r}_{12}$ corresponds to the interelectronic distance, while vectors $\bm{r}_A$ and $\bm{r}_B$ denote the intra-atomic radius vectors drawn from the nucleus to the bound electron of atoms $A$ and $B$, respectively. The vertices indicate the nuclear ($Z_A e$, $Z_B e$) and electronic ($-e_A$, $-e_B$) charges. The diagram allows one to extract the dependence of the photon propagator on vectors $\bm{R}$ and $\bm{r}_{AB}$. The former is used in the standard approach to define the van der Waals interaction, while $\bm{r}_{AB}$ is required for defining the magnetic moments of the electrons.}
\label{fig:2}
\end{figure}
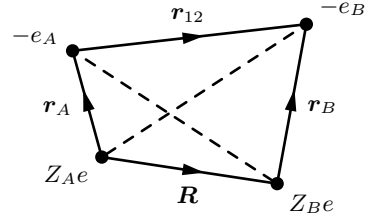

The standard procedure for subsequent calculations employs the adiabatic approximation and the substitution $r_{12} \rightarrow R$ in the photon propagator, where $R$ is the internuclear distance. While retaining the approximation of slowly varying nuclear positions, this substitution can be refined. To do so, the vector $\bm{r}_{12}$ should be expressed through $\bm{R}$ taking into account the electronic radius vectors, see Fig.~\ref{fig:2}. It is straightforward to see that $r_{12}=|\bm{R}-\bm{r}_{AB}|$, where $\bm{r}_{AB}=\bm{r}_A-\bm{r}_B$.

Further calculations can be performed analogously to classical electrodynamics for computing the field produced by a system of charges at distances significantly exceeding its size (such as that of atoms $A$ or $B$). For this purpose, we employ the Taylor series expansion of the function about a fixed internuclear distance:
\begin{eqnarray}
    \label{10}
    f(\bm{R}-\bm{r}_{AB})\approx f(\bm{R}) - (\bm{r}_{AB}\bm{\nabla}_{\bm{R}})f(\bm{R})
    \\
    \nonumber
    + \frac{1}{2}(\bm{r}_{AB}\bm{\nabla}_{\bm{R}})(\bm{r}_{AB}\bm{\nabla}_{\bm{R}})f(\bm{R})+\dots
\end{eqnarray}
Through a sequence of analytic steps, one can derive
\begin{eqnarray}
    \label{11}
    \frac{e^{\pm\ii |k_0| r_{12}}}{r_{12}}\approx \frac{e^{\pm\ii |k_0| R}}{R} + \frac{(\bm{r}_{AB}\bm{R})}{R}\left(\frac{1}{R}\mp\ii |k_0|\right)\frac{e^{\pm\ii |k_0| R}}{R}\,\,\,
    \nonumber
    \\
    + \frac{r^2_{AB}}{2}\left(\pm\frac{\ii |k_0|}{R} - \frac{1}{R^2}\right) \frac{e^{\pm\ii |k_0| R}}{R} \qquad
    \\
    \nonumber
    + \frac{(\bm{r}_{AB}\bm{R})(\bm{r}_{AB}\bm{R})}{2R^2}\left(-k_0^2\mp\frac{3\ii |k_0|}{R}+\frac{3}{R^2}\right) \frac{e^{\pm\ii |k_0| R}}{R}.
\end{eqnarray}

We now consider that the electronic 'motion' is much faster than the variation of the internuclear separation. This allows us to average the scalar product $(\bm{r}_{AB}\bm{R})$ over the angle between the vectors, see \cite{Landau} for the diamagnetic term. This averaging yields $\langle (\bm{r}_{AB} \bm{R}) \rangle = 0$ and $\langle (\bm{r}_{AB} \bm{R})^2 \rangle = (1/3)r^2_{AB}R^2$. These relations yield a significant simplification of expression (\ref{11}):
\begin{eqnarray}
    \label{12}
\frac{e^{\pm\ii |k_0| r_{12}}}{r_{12}}\approx \frac{e^{\pm\ii |k_0| R}}{R}-\frac{r_{AB}^2k_0^2}{6}\frac{e^{\pm\ii |k_0| R}}{R}.
\end{eqnarray}

Proceeding further requires one to consider the action of the gradient operators in formulas (\ref{8}) and (\ref{9}) on the functions in (\ref{12}). Here, the derivatives are understood to be taken with respect to the components of the vector $\bm{R}$. The result is
\begin{widetext}
\begin{eqnarray}
    \label{13}
\left(\delta_{ij}+\frac{\bm{\nabla}_{R_i}\bm{\nabla}_{R_j}}{k_0^2}\right)\left\lbrace -\frac{e^{\pm \ii |k_0| r_{12}}}{r_{12}}\right\rbrace\approx
\delta_{ij}\left(\frac{1}{k_0^2R^2}\mp \frac{\ii}{|k_0| R}-1\right) 
\frac{e^{\pm\ii |k_0| R}}{R}
+ \frac{R_i R_j}{R^2}\left(1\pm \frac{3\ii}{|k_0| R}-\frac{3}{k_0^2R^2}\right) 
\frac{e^{\pm\ii |k_0| R}}{R}\qquad
\nonumber
\\
+ \delta_{ij} \frac{r_{AB}^2}{6}\left(k_0^2 \pm\frac{\ii |k_0|}{R} - \frac{1}{R^2}\right) \frac{e^{\pm\ii |k_0| R}}{R}
+ \frac{R_i R_j}{R^2} \frac{r_{AB}^2}{6} \left(\frac{3}{R^2} \mp\frac{3 \ii |k_0|}{R} - k_0^2\right) 
\frac{e^{\pm\ii |k_0| R}}{R}.\qquad\qquad
\end{eqnarray}
\end{widetext}
This expression completely defines the photon propagators, following Eqs.~(\ref{8}) and (\ref{9}). 

The first line of Eq.~(\ref{13}) reproduces the well-known result, see, e.g., \cite{Berest,ZS-LR}. The second line can be viewed as a correction that is retained in the leading order of contributions from the one-photon exchange diagrams in Fig.~\ref{fig:1}. In order to demonstrate this, we begin by integrating over the time variables in Eq.~(\ref{1}). Then, after the integration over $k_0$ we apply the relation
\begin{eqnarray}
    \label{14}
    S_{AB}^{(2)} = -2\pi \ii \delta(\varepsilon^{A}_{a'}-\varepsilon^{A}_{a}+\varepsilon^{B}_{b'}-\varepsilon^{B}_{b}) U_{AB}^{(2)}(R).
\end{eqnarray}
The amplitude of the process $U_{AB}^{(2)}$ can be expressed as
\begin{eqnarray}
    \label{15}
    U_{AB}^{(2)}(R)=e^2\langle a b| \bm{\alpha}^{i}_A D_{ij}(\varepsilon_{aa'},R) \bm{\alpha}^{j}_B | a' b'\rangle.
\end{eqnarray}
By defining the transition energy as $\varepsilon_{aa'} \equiv \varepsilon^{A}_{a} - \varepsilon^{A}_{a'}$, the photon propagator can be expressed as the sum of zero-temperature and finite-temperature contributions:
\begin{equation}
    D_{ij}(\varepsilon_{aa'},R) = D_{ij}^{0}(\varepsilon_{aa'},R) + D_{ij}^{\beta}(\varepsilon_{aa'},R),
\end{equation}
corresponding to the terms defined in Eqs.~\eqref{8} and \eqref{9}. The subscripts $A$ and $B$ in the Dirac $\alpha$-matrices refer to the respective atoms. The matrix element $\langle a b|f(\bm{r}_A,\bm{r}_B)| a' b'\rangle$ is interpreted as $\langle \psi_a(\bm{r}_A)\psi_b(\bm{r}_B)|f(\bm{r}_A,\bm{r}_B)| \psi_{a'}(\bm{r}_A)\psi_{b'}(\bm{r}_B)\rangle$, following the conventional QED framework. The contributions from the direct and exchange diagrams, i) and ii) in Fig.~\ref{fig:1}, from Eq.~(\ref{15}) arise when the indices are set to $a'=a$, $b'=b$ and $a'=b$, $b'=a$, respectively.

Substituting equation~(\ref{13}) into (\ref{15}), one can find that for two identical atoms, the contribution that does not contain $r_{AB}^2$ vanishes in the non-relativistic limit. This is a consequence of the Dirac $\alpha$-matrix being equivalent to the momentum operator: $\psi^{+}\bm{\alpha}\psi \approx \phi^{+}\frac{\hat{\bm{p}}}{m}\phi$, where $\phi$ is the solution of Schr\"odinger equation and $\hat{\bm{p}}$ is the electron momentum operator. Then, via commutation relation $[\bm{r},\hat{H}]=\frac{\ii}{m}\bm{p}$ the electron dipole moment can be obtained. In turn, the diagonal matrix element of the dipole moment is zero \cite{LabKlim}. As a result, van der Waals interactions arise only in the fourth order in the electron–photon coupling.

Considering the second part of Eq.~(\ref{13}), it is possible to find the magnetic moment of a bound electron. The next higher multipoles should arise upon further Taylor series expansion in Eq.~(\ref{10}), see, e.g., \cite{Lopez-2025}. The expression for $U_{AB}^{(2)}$ can be transformed using the relation $r^2_{AB}=r^2_A+r_B^2-2\bm{r}_A\bm{r}_B$. We note immediately that the diagonal matrix elements proportional to $r_A^2$ or $r_B^2$ vanish for the same reasons discussed above (the one of $\alpha$-matrices yields a dipole moment in the non-relativistic limit). Thus, for the operator in the second line of Eq.~(\ref{13}), we have
\begin{eqnarray}
    \label{16}
\frac{e^2}{3} \frac{e^{\pm \ii |\varepsilon_{aa'}| R}}{R} \left[(\bm{\alpha}_A \bm{\alpha}_B)(\bm{r}_A \bm{r}_B) 
\left(\frac{1}{R^2} \mp\frac{\ii |\varepsilon_{aa'}|}{R} -\varepsilon_{aa'}^2 \right) 
\right.
\\
\nonumber
\left.
+ \frac{(\bm{\alpha}_A \bm{R})(\bm{\alpha}_B \bm{R})}{R^2}(\bm{r}_A \bm{r}_B) 
\left(\varepsilon_{aa'}^2 \pm\frac{3 \ii |\varepsilon_{aa'}|}{R} - \frac{3}{R^2} \right)\right].
\end{eqnarray}

The following transformations concern the extraction of the magnetic dipole interaction. This can be accomplished using the relations:
\begin{widetext}
\begin{eqnarray}
    \label{17}
(\bm{\alpha}_A \bm{\alpha}_B)(\bm{r}_A \bm{r}_B) = (\bm{\alpha}_A \bm{r}_B)(\bm{r}_A \bm{\alpha}_B)
+ ([\bm{r}_A\times \bm{\alpha}_A][\bm{r}_B\times \bm{\alpha}_B]),
\qquad\qquad
\\
\nonumber
\left(\bm{R}[\bm{r}_A\times \bm{\alpha}_A]\right)\left(\bm{R}[\bm{r}_B\times \bm{\alpha}_B]\right)  = R^2[\bm{r}_A\times \bm{\alpha}_A][\bm{r}_B\times \bm{\alpha}_B]
-\left[\bm{R}\times [\bm{r}_A\times \bm{\alpha}_A]\right]\left[\bm{R}\times [\bm{r}_B\times \bm{\alpha}_B]\right],
\end{eqnarray}
\end{widetext}
where the cross stands for the vector product. The second line in Eq.~(\ref{17}) can be recast in a standard form as a combination of dot products of the vectors involved. The introduction of the magnetic moment operator $\bm{\mu} = e[\bm{r} \times \bm{\alpha}]/2$ leads to an amplitude of the form:
\begin{eqnarray}
    \label{18}
e^{\pm \ii |\varepsilon_{aa'}| R}\langle a b| \left[(\bm{\mu}_A \bm{\mu}_B)
\left(\frac{1}{R^3} \mp\frac{\ii |\varepsilon_{aa'}|}{R^2} - \frac{\varepsilon_{aa'}^2}{R} \right) \qquad
\right.
\\
\nonumber
\left.
- \frac{(\bm{\mu}_A \bm{R})(\bm{\mu}_B \bm{R})}{R^2} 
\left(\frac{3}{R^3} \mp\frac{3\ii |\varepsilon_{aa'}|}{R^2} - \frac{\varepsilon_{aa'}^2}{R}\right)
\right] | a' b'\rangle + \text{rem}.
\end{eqnarray}

Under the condition $\varepsilon_{aa'}=0$ (unchanged atomic state of the two identical atoms) for simplicity, the remaining 'rem' terms in Eq.~(\ref{18}) can be written as 
\begin{eqnarray}
    \label{19}
\frac{e^2}{R^5}\left[\frac{5}{6}(\bm{\alpha}_A\bm{\alpha}_B)(\bm{r}_A\bm{r}_B)R^2 - \frac{1}{2}(\bm{\alpha}_A\bm{r}_B)(\bm{\alpha}_B\bm{r}_A)R^2 \qquad
\right.
\nonumber
\\
\left.
- \frac{7}{4}(\bm{\alpha}_A\bm{R})(\bm{\alpha}_B\bm{R})(\bm{r}_A\bm{r}_B) 
-\frac{3}{4}(\bm{\alpha}_A\bm{\alpha}_B)(\bm{r}_A\bm{R})(\bm{r}_B\bm{R})
\right. \,\,\,
\\
\nonumber
\left.
+\frac{3}{4}(\bm{\alpha}_A\bm{R})(\bm{\alpha}_B\bm{r}_A)(\bm{r}_B\bm{R})
+\frac{3}{4}(\bm{\alpha}_A\bm{r}_B)(\bm{\alpha}_B\bm{R})(\bm{r}_A\bm{R})
\right].
\end{eqnarray}
As expected (see, e.g., Appendix A in \cite{Lopez-2025}), this contribution can be identified with the quadrupole interaction. However, to construct the complete exchange interaction involving the quadrupole momenta $r_{A}^2\delta_{ij}+3r_{Ai}r_{Aj}$ and $r_{B}^2\delta_{ij}+3r_{Bi}r_{Bj}$ in combination with the projector $\delta_{kl}+R_kR_l/R^2$, it is necessary to consider the next-order terms in the expansion (\ref{10}). According to the selection rules, the angular momentum of the state should change by two units in this case. Consequently, within the scope of our study, the terms in Eq.~(\ref{19}) can be neglected, just as was done earlier for the $r_A^2$ and $r_B^2$ terms.


Thus, for two identical atoms in the same states expression (\ref{18}) in the limit $\varepsilon_{aa'}=0$ yields the magnetic interaction energy as
\begin{eqnarray}
    \label{21}
\Delta E_{AB}^{(m)} = \frac{\langle (\bm{\mu}_A\bm{\mu}_B)\rangle}{R^3} 
- \frac{3\langle (\bm{\mu}_A \bm{R}) (\bm{\mu}_B \bm{R})\rangle}{R^5},
\end{eqnarray}
where averaging over the two-atom state $|a b \rangle$, as for the diagonal matrix element in Eq.~(\ref{15}), is implied. The resulting Eq.~(\ref{21}) is well known from the phenomenological generalization of the classical result to the quantum case. We, however, have derived it within the framework of a rigorous quantum electrodynamics theory.

Turning now to the case of finite temperatures, it should be noted that in the limit $\varepsilon_{aa'}=0$ the Planck distribution function diverges. However, since the thermal photon propagator (\ref{9}) contains the difference of exponents, the final expression is finite,  giving an additional factor of $-2\ii R/\beta$. Accounting to Eq.~(\ref{9}), this yields 
\begin{eqnarray}
    \label{22}
\Delta E_{AB}^{(m)\beta} = -\frac{2\ii}{\beta}\left[\frac{\langle (\bm{\mu}_A\bm{\mu}_B)\rangle}{R^2} - \frac{3\langle (\bm{\mu}_A \bm{R}) (\bm{\mu}_B \bm{R})\rangle}{R^4}\right].
\end{eqnarray}

From expressions (\ref{21}) and (\ref{22}), it follows that the sign of the magnetic interaction for pairs such as hydrogen-antihydrogen (with oppositely charged constituent particles) is opposite to that for pairs of ordinary atoms, at both zero and finite temperature. The charge introduced in the initial expression (\ref{1}) was implicitly defined for the electron. For the positron, one would accordingly write $+\ii e$. In formulas (\ref{21}) and (\ref{22}), this convention is reflected in the opposite sign of the Bohr magneton in the non-relativistic limit.

\section{Long-distance coupling through hyperfine states}
\label{hfs}
In the previous section, we were interested in the long-range magnetic interaction of two atoms without changing their internal states. The interaction potential for two magnetic dipoles, well known from classical electrodynamics, was derived within the QED formalism by describing the one-photon exchange between the atoms. In this approach, the coupling potential (\ref{21}) emerges under the assumption $\varepsilon_{aa'}= 0$ and allows for a straightforward generalization to the case of interaction stimulated by external thermal radiation (\ref{22}), see also Eq.~(\ref{9}).

Expression (\ref{18}), nevertheless, allows one to rather straightforwardly transition to the description of magnetic interatomic interaction taking into account the hyperfine structure (HFS) of the atomic energy levels. In this case, the energy difference between levels $a$ and $a'$ ($b$ and $b'$) can be set equal to the hyperfine splitting $\varepsilon_{aa'}=\Delta_{\rm HFS}\neq 0$. This circumstance is due to the fact that the interaction operator, given by expression (\ref{18}), does not change the parity of the states. Furthermore, because the HFS is small, such a process is the least energy-intensive (compared, for example, to the fine structure).

Separating into real and imaginary parts, for the zero-temperature contribution we find
\begin{widetext}
\begin{eqnarray}
    \label{hfs.1}
\Re\Delta E_{\rm HFS}^{(m)} =\langle ab|(\bm{\mu}_A\bm{\mu}_B)|a'b'\rangle 
\left(\frac{1}{R^3}\cos{(\Delta_{\rm HFS} R)} +
\frac{|\Delta_{\rm HFS}|}{R^2}\sin{(|\Delta_{\rm HFS}|R)} - \frac{\Delta_{\rm HFS}^2}{R}\cos{(\Delta_{\rm HFS} R)}\right)
\\
\nonumber
+\frac{\langle ab|(\bm{\mu}_A\bm{R})(\bm{\mu}_B\bm{R}))|a'b'\rangle}{R^2}
\left(\frac{\Delta_{\rm HFS}^2}{R}\cos{(\Delta_{\rm HFS} R)} - \frac{3 |\Delta_{\rm HFS}|}{R^2} \sin{(|\Delta_{\rm HFS}| R)} - \frac{3}{R^3}\cos{(\Delta_{\rm HFS} R)}\right),
\\
\nonumber
\Im\Delta E_{AB}^{(m)} = \langle ab|(\bm{\mu}_A\bm{\mu}_B)|a'b'\rangle 
\left(\frac{1}{R^3}\sin{(|\Delta_{\rm HFS}| R)} -
\frac{|\Delta_{\rm HFS}|}{R^2}\cos{(\Delta_{\rm HFS} R)} - \frac{\Delta_{\rm HFS}^2}{R}\sin{(|\Delta_{\rm HFS}| R)}\right)
\\
\nonumber
+\frac{\langle ab|(\bm{\mu}_A\bm{R})(\bm{\mu}_B\bm{R}))|a'b'\rangle}{R^2}
\left(\frac{\Delta_{\rm HFS}^2}{R}\sin{(|\Delta_{\rm HFS}| R)} + \frac{3 |\Delta_{\rm HFS}|}{R^2} \cos{(\Delta_{\rm HFS} R)} - \frac{3}{R^3}\sin{(|\Delta_{\rm HFS}| R)}\right).
\end{eqnarray}
\end{widetext}
Here, $\Re$ and $\Im$ denote the real and imaginary parts, respectively. The modulus sign (omitted where it is not essential) indicates that the energy difference is always taken to be positive.

In turn, for the interaction stimulated by thermal radiation, Eq. (\ref{18}) yields
\begin{widetext}
\begin{eqnarray}
    \label{hfs.2}
\Im\Delta E_{\rm HFS}^{(m)\beta} = \langle ab|(\bm{\mu}_A\bm{\mu}_B)|a'b'\rangle 
\left(\frac{2}{R^3}\sin{ (|\Delta_{\rm HFS}| R)} - \frac{2 |\Delta_{\rm HFS}|}{R^2}\cos{(\Delta_{\rm HFS} R)} - \frac{2 \Delta_{\rm HFS}^2}{R}\sin{(|\Delta_{\rm HFS}| R)}\right)n_\beta(|\Delta_{\rm HFS}|)
\\
\nonumber
+\frac{\langle ab|(\bm{\mu}_A\bm{R})(\bm{\mu}_B\bm{R}))|a'b'\rangle}{R^2}
\left(\frac{2 \Delta_{\rm HFS}^2}{R}\sin{(|\Delta_{\rm HFS}| R)} + \frac{6 |\Delta_{\rm HFS}|}{R^2}\cos{(\Delta_{\rm HFS} R)} - \frac{6}{R^3}\sin{ (|\Delta_{\rm HFS}| R)}\right)n_\beta(|\Delta_{\rm HFS}|).
\end{eqnarray}
\end{widetext}

The energy conservation law, enforced by the $\delta$-function $\delta(\varepsilon^{A}_a-\varepsilon^{A}_{a'}+\varepsilon^{B}_b-\varepsilon^{B}_{b'})$ in the $S$-matrix element, allows two channels for magnetic interaction: i) without change in the atomic states, i.e., $\varepsilon^{A}_a=\varepsilon^{A}_{a'}$ and $\varepsilon^{B}_b=\varepsilon^{B}_{b'}$ (discussed above), and ii) where the de-excitation of one atom results in the excitation of the other. The following description refers to the second case.

For two identical one-electron atoms, using hydrogen as an example, the hyperfine structure is accounted for by coupling the total angular momentum with the nuclear spin via Clebsch-Gordan coefficients. Then, after carrying out the necessary angular algebra \cite{VMK} (i.e., averaging over the projections of the initial state and summing over those of the final state of the total atomic angular momentum), for the simplest case of $s$-states, the expressions simplify to
\begin{eqnarray}
    \label{hfs.3}
\Re\Delta E_{\rm HFS}^{(m)} = 
-\frac{2 \Delta_{\rm HFS}^2 \mu_0^2}{9 R}  \cos{(\Delta_{\rm HFS} R)},
\\
\nonumber
\Im\Delta E_{\rm HFS}^{(m)} = - \frac{2 \Delta_{\rm HFS}^2 \mu_0^2 }{9 R}\sin{(|\Delta_{\rm HFS}| R)},
\end{eqnarray}
with $\mu_0$ being the electron Bohr magneton. The thermally induced contribution is given by
\begin{eqnarray}
    \label{hfs.4}
\Im\Delta E_{\rm HFS}^{(m)\beta} = 
-\frac{4\Delta_{\rm HFS}^2 \mu_0^2}{9 R} \sin{(|\Delta_{\rm HFS}| R)}n_\beta(|\Delta_{\rm HFS}|).
\end{eqnarray}

\section{Results and discussion}
\label{RD}
The calculations presented in Sec.~\ref{section1} were performed for identical states of two one-electron atoms (considering diagonal matrix elements). Nevertheless, expressions (\ref{16}) and (\ref{18}) admit generalization to an arbitrary case. This requires taking into account that $\varepsilon_{aa'}\neq 0$. Building on this, we have examined the interaction of two atoms with hyperfine structure in Sec.~\ref{hfs}. The restriction to hyperfine structure stems from the fact that the magnetic interaction operator obtained does not change state parity. In a many-electron atom, the states $a, a', b, b'$ are then to be interpreted as the overall state with a definite total angular momentum and multiplicity. 

The preceding sections were likewise confined to a description of the 'direct' one-photon exchange. However, it is necessary to include the 'exchange' diagram, depicted in Fig.~\ref{fig:1} ii). Considering that the total state of the two equivalent atoms can be either symmetric (plus sign in Fig.~\ref{fig:1}) or antisymmetric (minus sign in Fig.~\ref{fig:1}), expressions (\ref{hfs.3}), (\ref{hfs.4}) will either double or vanish, respectively.

The numerical evaluation of formulas (\ref{21}), (\ref{22}) and (\ref{hfs.3}), (\ref{hfs.4}) pertains to the calculation of matrix elements of the magnetic dipole moments. In the non-relativistic limit, the operator $\bm{\mu}$ defined above via the Dirac $\bm{\alpha}$-matrix reduces to $\bm{\mu} = \mu_0(\bm{l} + 2\bm{s})$, where $\bm{l}$ is the electron's orbital angular momentum operator and $\bm{s}$ is the corresponding spin operator ($\mu_0 = e\hbar/(2m c)$ is the Bohr magneton). Note also that in the non-relativistic limit, only matrix elements with coinciding principal quantum numbers are non-zero. In the opposite case, relativistic corrections to the operator and the wave functions should be taken into account \cite{Sucher_md}. By performing the necessary angular algebra \cite{VMK} and averaging over the initial state projections while summing over the final ones, it can be shown that for $s$-states
\begin{eqnarray}
    \label{23}
\Delta E_{AB}^{(m)} = \frac{\mu_0^2}{R^3},
\\
\nonumber
\Delta E_{AB}^{(m)\beta} = -\frac{2\ii}{\beta}\frac{\mu_0^2}{R^2},
\end{eqnarray}
and
\begin{eqnarray}
    \label{24}
\Delta E_{\rm HFS}^{(m)} &=& 
-\frac{\Delta_{\rm HFS}^2 \mu_0^2}{3 R}  e^{\ii\Delta_{\rm HFS} R},
\\
\nonumber
\Delta E_{\rm HFS}^{(m)\beta} &=& 
-\ii\frac{2\Delta_{\rm HFS}^2 \mu_0^2}{3 R} \sin{(|\Delta_{\rm HFS}| R)}n_\beta(|\Delta_{\rm HFS}|).
\end{eqnarray}
The resulting expressions, written in terms of the electron Bohr magneton $\mu_0$ for hydrogen-like atoms. 

Taking the Bohr magneton in atomic units, $\mu_0 = 1/2$, the interaction energy of Eq.~(\ref{23}) for atoms $A$ and $B$ in $s$-states evaluates to $\Delta E_{AB}^{(m)} \approx 1.331\times 10^{-5}/R^3$ a.u. At an internuclear separation of $R = 10$ a.u., this yields approximately $87.6$ MHz. At the equilibrium internuclear distance of the hydrogen molecule, $R \approx 1.4$ a.u., the contribution reaches the order of $10^{-6}$ a.u., which is consistent with the known magnitude of the spin-spin relativistic correction in H$_2$. The thermally induced contribution, evaluated using $k_{\rm B} = 3.16681\times 10^{-6}$ a.u., gives $\Delta E_{AB}^{(m)\beta} \approx -1.85\times 10^{-10}\ii/R^2$ a.u. at $T = 300$ K, which translates into a purely imaginary part approximately equal to ${\rm Im}\Delta E_{\rm HFS}^{(m)\beta} = 2.43\times10^{4}$ s$^{-1}$ at $R = 10$ a.u.

For the interaction channel involving a possible transition between hyperfine sub-levels, see Eq.~(\ref{24}), one can readily obtain the following. For the $1s$ state in the hydrogen atom, with $\Delta_{\rm HFS}^{(1s)}\approx 2.15878\times 10^{-7}$ a.u., the dispersion interaction coefficient in atomic units is given by $-1.10\times 10^{-23}[\cos{(\Delta_{\rm HFS}R)}+\ii\sin{(\Delta_{\rm HFS}R)}]/R$. Here we can approximate the expression in square brackets by unity. This contribution is negligible in contrast to the non-perturbed states and, at the same time, shows a change in the scaling with the interatomic distance from $1/R^3$ to $1/R$. In turn, the interaction induced by thermal radiation is given by a coefficient twice as large as the previous one. However, there is an additional factor represented by the Planck distribution function at the resonant frequency. For hyperfine structure, this factor plays the role of an enhancement. At room temperature for the $1s$ state, $n_\beta(|\Delta_{\rm HFS}|) = 4401.6$, leading to a dispersion coefficient (purely imaginary) of $-9.7\times10^{-20}\sin(\Delta_{\rm HFS}R)\, \ii/R$ a.u. Thus, including the hyperfine structure makes the induced imaginary part of the correction four orders of magnitude larger than its natural counterpart, yet it remains small at the present level of experimental precision.

It should be emphasized that the nature of the contributions in (\ref{24}) is fundamentally different from that of the magnetic dipole–dipole interaction. Whereas the latter represents a well‑known relativistic correction to the interaction energy of atoms in unperturbed states, the present effect arises from transitions between hyperfine sublevels, thereby accounting for the internal atomic structure, and is therefore an inherently quantum electrodynamic phenomenon, with no classical analogue in the static spin‑spin interaction. This situation is analogous to that discussed in \cite{Jentschura2023}, where the resonant one‑photon dipole exchange between identical atoms - one in the ground state and one in an excited state - gives rise to a qualitatively similar long‑range behavior of the interaction. Another manifestation of QED theory is that the long‑range interaction involves precisely the magnetic moment of the atom, which in principle also includes the nuclear magnetic moment (though not considered in this work).

Another application of the long-range magnetic interaction under investigation, which is no less significant, is as follows. One of the most accurately measured quantities is the frequency of the $1s-2s$ transition in the hydrogen atom, known with an absolute error of about $10$ Hz  \cite{Mohr-2016-RMP}. 
Because the found energy interaction is the same for $ns$-states with different principal quantum numbers, it is expected to cancel out almost completely (to within relativistic and QED corrections). However, accounting for the 'exchange' contribution to the symmetric and antisymmetric diatomic states can yield a non-zero result. The fractions of initial and final symmetric or antisymmetric diatomic states become thus relevant in atomic-beam or magneto-optical trap measurements. As is known from the theory of long-range dipole-dipole interactions, reducing the atomic density attenuates this effect. Increasing the interparticle distance $R$ from $10$ a.u. to $1$ $\mu$m renders such long-range magnetic interaction negligible, reducing it by a factor of approximately $10^{-10}$ and inducing a shift comparable to the absolute frequency uncertainty of the $1s-2s$ transition at an interatomic separation of approximately $0.1\mu$m. Nevertheless, in studies of matter–antimatter interactions, e.g., in hydrogen–antihydrogen systems, these effects (including the sign reversal relative to the hydrogen molecule) can play a decisive role.

\section{Conclusions}
\label{concl}
To conclude the derivations presented above, several remarks are in order.
First, for the ground state of the system, the expressions simplify significantly under the assumption that the individual atoms remain in their initial states, $\varepsilon_{aa'} = 0$.
It should be noted that the thermally induced interaction exhibits a distance dependence that falls off one inverse power more slowly.
Nonetheless, this magnetic long-range interaction is further suppressed by a factor of $\alpha^2$ (where $\alpha$ is the fine-structure constant) due to the temperature-dependent coefficient.
Furthermore, the zero-temperature and finite-temperature magnetic dipole interactions differ for the symmetric and antisymmetric states of a system of identical atoms~\cite{Adhikari:2017,Adhikari:2017:2}. If, however, different atoms are considered, only the one 'direct' Feynman diagram shown in Fig. 1 i) needs to be taken into account. 
Then, the difference in the magnetic long-range interaction between hydrogen atoms and between a hydrogen atom and an antihydrogen atom becomes immediately apparent. 
Another important point is that the one-photon exchange yields a well-defined asymptotic form, which is frequency-independent, unlike the two-photon dipole-dipole long-range potential.

Lastly, attention should be paid to the contributions represented by the imaginary parts of expressions (\ref{23}) and (\ref{24}). Two of them are negative (see Eq.~(\ref{24})), while one is positive (the second line in Eq.~(\ref{23})) and dominant. Relating the obtained contributions to the imaginary part of the atom–atom magnetic polarizability, negative contributions are quite familiar. In atomic physics, they are associated with the decay of the system. The positive imaginary part, in turn, corresponds to standard van der Waals attraction with dissipation. Two hydrogen atoms interacting via the thermal field will tend toward the minimum‑energy configuration, with the excess energy being dissipated as heat. In the case of the hydrogen–antihydrogen system, the sign flips to the opposite, leading to attraction and subsequent annihilation. Thus, the thermal imaginary part in (\ref{23}) can be interpreted as a thermally induced decay channel. Consequently, for the hydrogen–antihydrogen system, the imaginary contribution in Eq.~(\ref{23}) directly yields the annihilation rate induced by blackbody radiation. A key implication is that higher temperatures stimulate annihilation between hydrogen and antihydrogen atoms. This conclusion appears plausible for explaining the present‑day absence of antimatter atoms in the Universe, given that annihilation is enhanced at higher temperatures according to the established linear dependence of the long‑range magnetic interaction. We restrict ourselves to this qualitative explanation, leaving a more detailed study of this intriguing question concerning the imaginary part of the interaction potential for future work \cite{Dicke_narrowing,Volokitin:2007,PhysRevA.76.013818,Jentschura2023,PhysRevLett.134.133603}.



\section*{Acknowledgements}
This work was supported by the Russian Science Foundation under grant \textnumero{25-22-00625}.
\bibliography{mybibfile}

\end{document}